\documentclass{aa}
\usepackage{graphicx}

\begin{document}
\def\etal{{\it et al.}}
\def\lae{\mathrel{<\kern-1.0em\lower0.9ex\hbox{$\sim$}}}
\def\gae{\mathrel{>\kern-1.0em\lower0.9ex\hbox{$\sim$}}}

\title {VLBA observations of GHz-Peaked-Spectrum \\ radio sources at 15 GHz}

\author{
C. Stanghellini\inst{1,6} \and D. Dallacasa\inst{2} 
\and C.P. O'Dea\inst{3} \and S.A. Baum\inst{3}
\and R. Fanti\inst{4,5}
\and C. Fanti\inst{4,5}
}

\offprints{C. Stanghellini}

\institute{
Istituto di Radioastronomia del CNR, C.P. 141, I-96017 Noto SR, Italy 
\and 
Dipartimento di Astronomia, Universit\`a degli Studi, via Ranzani 1, I-40127
Bologna,  Italy
\and
Space Telescope Science Institute, 3700 San Martin Drive, Baltimore, MD, 21218 
\and
Dipartimento di Fisica, Universit\`a degli Studi, via Irnerio 46, I-40126 Bologna, 
Italy
\and
Istituto di Radioastronomia del CNR, Via Gobetti 101, I-40129 Bologna, Italy 
\and
Visitor at the Space Telescope Science Institute, Baltimore
}

\date{\today}

\abstract{We present VLBA observations at 15 GHz
of ten GHz Peaked Spectrum (GPS) radio sources.
The cores are often difficult or impossible
to locate. When likely cores are found, they
account for a small fraction 
of the  flux density in GPS galaxies - around or below 2\%, 
while in GPS quasars they can account for more than 20\%
of the total flux density.
We detect low polarization in the GPS sources --
i.e., typically less than a few percent and often 
less than one percent. This establishes that low polarization
in the parsec scale structure is an important defining
characteristic of the GPS sources. 
The dichotomy in the radio morphology versus optical 
identification, i.e., galaxies are  symmetric and quasars are not,
is basically confirmed from these new data, which
also indicate that the radio emission from GPS
quasars is dominated by a jet, with often a weak or
hidden core, suggesting they are
at moderate angles to the line of sight, and so are only
moderately beamed. 
}
\maketitle
\section{Introduction}

A small but still significant fraction of
compact radio sources (i.e., radio sources
strongly dominated by an unresolved component
on arc-second scales) show a convex radio spectrum
peaking around 1 GHz.

These sources have caught the interest of astronomers 
because they have remarkable properties. They are a mixed 
group of galaxies and quasars, with the quasars
usually found at very high redshifts, and
with the galaxies showing a parsec scale morphology
resembling that of the large-scale ``classical doubles" which 
span hundreds of kpc (e.g., Phillips \& Mutel 1982, Mutel,
Hodges, \& Phillips 1985; Conway et al. 1994; Readhead et al. 
1996a; Taylor, Readhead, \& Pearson 1996; Stanghellini et al. 
1997, 1999). These small classical doubles
were named ``Compact Symmetric Objects" (CSOs) by
Wilkinson et al. (1994).  

This led to study the GPS radio sources because they are likely to be
the progenitors of powerful radio galaxies, and if so,
can shed light on the onset of the radio source phenomenon.
In addition, the  study of the subclass of GPS quasars 
has the goal of searching for high redshift AGN with which to  
probe the environment of galaxies in the early Universe.

Owsianik and Conway (1998) and Owsianik et al. (1998) were the first
to report on direct evidence of proper motion of the outer edges in CSOs, 
with an
estimated dynamical age of the order of $10^3$ years, and a strong
upper limit of $10^4$ years on their actual age, thus providing
strong support to the hypothesis that these sources are
the youngest phases of the lobe dominated radio sources.
Nowadays, expansion velocities of the outer edges of radio source 
have been detected or suspected in about ten CSOs (Fanti 2000).

A comprehensive review on GPS radio sources
and our current understanding of their nature
has been published by O'Dea (1998).

Over the past decade,  we have concentrated our efforts on studying 
in detail the brightest members of this class of radio sources, selecting
a complete flux-limited sample, and 
extensively observing it in radio, optical, and X-ray bands.

The complete sample (Stanghellini et al. 1998)
consists of 33 objects, with the following 
selection criteria: 

- declination $\delta > -25^o$

- galactic latitude $|l| > 10^\circ$

- flux density at 5 GHz $S_{5GHz}>1Jy$

- turnover frequency between 0.4 and 6 GHz

- spectral index $\alpha _{thin} < -0.5$ ($S_{\nu}\propto \nu^{\alpha}$)
in the high frequency, likely optically thin, part of the spectrum. 

Much can be learned from the radio morphology of these sources 
as revealed by VLBI observations.

Stanghellini et al. (1997, paper I) presented global VLBI  
5 GHz images of 9 radio sources of the GPS 1 Jy sample. 
Stanghellini et al. (1999) presented 5 GHz images
of additional 11 radio sources from the list of O'Dea et al. (1991), 
including two objects from the GPS 1 Jy sample.

The results from these observations complemented with information 
from the literature for the sources of the complete sample
suggest that GPS quasars have preferentially core-jet 
or complex morphology while GPS galaxies are likely to be CSOs,
(with a few showing a complex morphology). 

Continuing our study of the milliarcsecond morphology
of our complete sample, we now present new polarization sensitive
VLBA 15 GHz images of several sources already observed at 5 GHz.
The new data allow us  to study the spectral properties
of the radio components which is necessary to locate the core
and unambiguously define the radio morphology.

Observations of the integrated polarization of GPS sources
with the VLA have shown that the fractional polarization is very low --
typically less than one percent (e.g., Rudnick \& Jones 1982;
Rusk 1988; O'Dea et al. 1990a; O'Dea, Baum \& Stanghellini 1991). In
this paper we address whether this is due to beam depolarization
in the VLA observations or if this is a genuine property
of the parsec scale structure. 

H$_0$=100 km sec$^{-1}$ Mpc$^{-1}$, and q$_0$=0.5 have been used
throughout this paper, and the spectral index $\alpha$
is defined such that $S_{\nu}\propto \nu^{\alpha}$.

\section{Observations and results}

The observations were carried out on 22 April 1996, at 15.365 GHz,
with the VLBA with the addition of a single VLA antenna, for a total
of eleven 25 meter antennae, recording 16 frequency channels in
2 baseband converters, for a total of 16 MHz bandwidth in each hand of
circular polarization.  Each source has been observed
for about 2 hours, with 3 or 4 scans at different hour angles
to improve the UV coverage.

After correlation in Socorro, the data
have been processed and analyzed with the Astronomical Image Processing
System (AIPS) developed by the National Radio Astronomy Observatory
(NRAO). The data were fringe fitted, calibrated, edited, 
self-calibrated, and imaged in a standard way.
Phase self-calibration has been applied several times, 
and a final iteration of amplitude and phase self-cal has been performed. 
The gain correction was only a few percent for all the sources
confirming the well known reliability of the a-priori calibration done with 
the antenna gains and system temperatures. 

In Fig. 1 we show the best (0738+313) and poorest (1143-245)
UV coverage in our data.

\begin{figure*}
      \includegraphics[width=18cm]{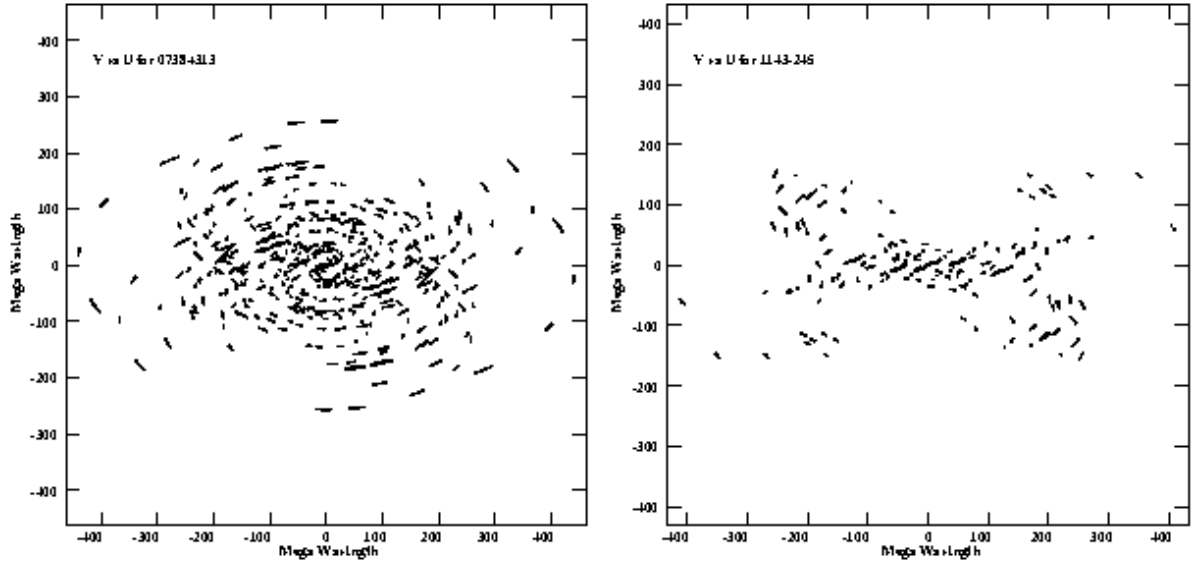}
\caption[]{the best and the poorest UV coverage}
\label{}
\end{figure*}

The list of the observed radio sources with basic information
is shown in Table 1. The typical noise on the cleaned images is around
 a half mJy and the typical restoring beam is of the order of 1 mas. 

Because the small size of several of the sources observed, 
we also show deconvolved images obtained using the Maximum Entropy Method 
(MEM), which increases the resolution for high signal to noise components.
These MEM images should be considered with caution; we use the
finer details they show to confirm what is seen
in the cleaned images, or to make suggestions to be confirmed by
future observations.

\begin{table*}
\caption{ GPS sources. Columns 1
through 12 provide: (1) name, (2) RA and (3) DEC (B1950 coordinates), (4)
optical identification, (5) optical magnitude and filter (6) redshift, 
(7) linear scale factor in pc/mas
[H$_o$=100 km s$^{-1}$ and q$_o$= 0.5 have been assumed]. A value of
$z=1$ has been adopted for the one Stellar Object with unknown redshift, 
(8) total size in pc, (9) flux density at 15 GHz, 
as measured  from VLBI images presented here, (10) low frequency spectral index
(11) high frequency spectral index, and (12) turnover
frequency (the values in column 10-12 are from
Stanghellini et al. 1998).
} 
\begin{flushleft}
\begin{tabular}{lcclllccrccr}
\hline
\hline
\noalign{\smallskip}
$Source$ & $RA$ & $DEC$ & $id$ & $m$ & $z$ & $scale$  & $l$ &
$S_{vlbi}$ &
$ \alpha_l$ & $\alpha_h$ & $\nu_m~ $ \\
 &$B1950$&$B1950$& & & & $_{pc/mas}$ & $pc$ &$Jy$& & 
&$_{GHz}$ \\
 (1)&(2)&(3)&(4) &(5) &(6) &(7)&(8)&(9)&(10) &(11) 
&(12) \\
\noalign{\smallskip}
\hline
\noalign{\smallskip}
 0500+019&05 00 45.18&+01 58 53.8&  G     &21.0i& 0.583&3.8&57&0.83 
&1.6 & -0.9 & 2.0\\
 0738+313&07 38 00.18&+31 19 02.1&  Q     &16.1V& 0.630&3.9&40&2.11 
&0.7 & -0.8 & 5.1\\
 0742+103&07 42 48.47&+10 18 32.6&  Q?    &23R& (1)  &4.3&(56)&1.31 
&0.7& -0.7& 2.8\\
 0743$-$006&07 43 21.05&$-$00 36 55.8&  Q     &17.5V& 0.994&4.3&26&1.28 
&0.88 & -0.61 & 5.8\\
 1143$-$245&11 43 36.37&$-$24 30 52.9&  Q     &18.5V& 1.95 &4.1&30&0.57 
&2.5 & -0.7 & 2.0\\
 1345+125&13 45 06.17&+12 32 20.3&  G     &15.5r& 0.122&1.5&130&1.05 
&0.9 & -0.7 & 0.6\\
 1404+286&14 04 45.62&+28 41 29.2&Sy&14.6r&0.077&1.0&7&0.98&1.5 
&-1.6&5.3\\
 2126$-$158&21 26 26.78&$-$15 51 50.4&  Q     &17.3V& 3.270&3.5&20&0.81 
&1.3 & -0.5 & 3.9\\
 2128+048&21 28 02.61&+04 49 04.4&  G     &23.3r& 0.99 &4.3&150&0.51 
&1.0 & -0.8 & 0.8\\
 2134+004&21 34 05.21&+00 28 25.1&Q&18.0V&1.936&4.1&12&5.23&1.2 
&-0.7&5.2\\
\noalign{\smallskip}
\hline 
\end{tabular}
\end{flushleft}
\end{table*}

The parameters of the various components are shown in Table 2 as
measured on the images. 
We report the Full Width Half Maximum (FWHM) size
derived from Gaussian fits (task JMFIT in AIPS).
In some cases the fit is rather poor and the
reported numbers, preceded by the symbol $\sim$, 
must be considered just a rough
estimate, otherwise we assume a typical 10$\%$ uncertainty for the diameters
and flux densities of the components. 

\begin{table*}
\caption{ Component parameters derived from the
15 GHz VLBA images.
Columns 1 through 13 give: (1) source name and our classification
(CSO:Compact Symmetric Object; CJ: core-jet; CX: complex), 
(2) maximum size in mas and (3) in parsec,
(4) component identification, (5) major and (6)  minor axes (fitted Gaussian FWHM 
for compact components, approximate detected size for extended components), 
(7) and (8) linear size
derived from columns 3 and 4, (9) position angle, (10)
total flux density of the
component, (11) intrinsic brightness temperature neglecting 
relativistic effects, (12) energy density adopting minimum energy, 
(13) equipartition magnetic field. The scale factor for each source is 
that given in column 5, Table 1. The component labels follow the order 
used in Paper I. 
\hfill\break
} 
\begin{flushleft}
\begin{tabular}{ccclcccccccccccc}
\hline
\hline
\noalign{\smallskip}
Source and & $\Theta$&L&Comp. & $\theta_1 $ & $\theta_2 $ &$l_1$ & $l_2$ & PA &
$S_{15GHz}$& T$_b$ &u$_{min}$ & H$_{eq}$     \\
class    &$mas$&pc&  & $mas$       & $mas$     & pc~    & pc~   & &   mJy    & 
$10^9K^\circ$   &$10^{-6}erg/cm^3$& $10^{-3}G$        \\ 
\noalign{\smallskip}
\hline 
\noalign{\smallskip}
 0500+019&13&50 & A1 & 0.9 & 0.6 &3.4&2.3 & +215 & 215 & 1.4 &400  &66     \\
     CSO && & A2 & 1.6 & 0.6 &6.1&2.3&  +0 & 25 &.09 & 84 &30     \\
         && & B1 & 0.8 & 0.3 &3.0&1.1& +164 & 446 &6.7  &1400&120   \\
         && & B2 & 0.8 & 0.4 &3.0&1.5& +14 & 56 &0.63 &320&58   \\
         && & C & 0.7 & 0.6 &2.7&2.3& +113 &  13 &0.11 &93 &32      \\
         && & D & 1.2 & 0.6 &4.6&2.3& +12 & 35 &0.17 &120&36     \\
 0738+313&5&20 & A & 0.6 & 0.1 &2.3 &0.4 & +5   & 925 &57 &9500&320    \\
      CX && & B & 0.6 & 0.3 &2.3&1.2& +155 &1061 &22 &2900&180   \\
 0742+103&4&17 & A1 & 1.3 & 0.4 &5.6&1.7& +150 &990 &8.6&1900&140     \\
  J&&& A2  & $\sim$1.2  & $\sim$1.2&$\sim$5.2&$\sim$5.2&-&$\sim$320&1.1&330&60  \\
 0743-006 &2&9& A1 & 0.4 & 0.3 &1.7&1.3& +38  &229 &8.6   &2200&160  \\
      CJ  &&& A2 & 0.4 & 0.2 &1.7&0.9& +46  &753 &42&7000&270  \\
          &&& A3 & 0.3 & 0.2 &1.3&0.9& +38  &281 &21&4700&220  \\
 1143-245 &5&20& A&1.0&0.6&4.1&2.5& +18   &460&5.1&2100&150 \\
       CJ &&& B & 2.2 & 0.6 &9&2.5&  -6  &87  &0.4&520  &75  \\
 1345+125 &75&110& A & 1.3 & 0.4 &2.0&0.6& +163 & 304 &1.5&520 &75  \\
       CSO&&& B & 2.3 & 0.9 &3.5&1.4& +143 & 259 &0.32&140 &38  \\
          &&& C & 4.0 & 1.4 &6.0&2.1& +167 & 66 &0.03&27&17  \\
 1404+286 &8&8& A & 0.5 & 0.3 &0.5&0.3& +162 & 865 & 14&2600 &170  \\
     CSO&&& C & $\sim 1$ & $\sim 1$ &$\sim 1$&$\sim 1$& - & 24 &.06  &57 &25  \\
        &&& D & $\sim 0.5$ & $\sim 0.5$&$\sim 0.5$&$\sim 0.5$& -& 7&.07&92&32\\
 2126-158 &3&10&A&$<$0.3&$<$0.3&$<$1.1&$<$1.1 &   -  & 690&$>$74&$>$28000&$>$550 \\
       CJ &&& B & 1.8 & 0.3 & 6.3& 1.1&  +8 & 110 & 2.0 &3500&200  \\
 2128+048 &35&150& A &$<$0.3&$<$0.3&$<$1.3&$<$1.3& - &  7.4 &$>$0.35&$>$360&$>$62  \\
      CSO &&& B & 0.8 & 0.2 &3.4& 0.9& +154 & 54 &1.5&1000&110  \\
    &&&C+D$_{ext}$& 4   &   4 &17&17&  - & 420 &0.12&44&22  \\
          &&& E & 1.0 & 0.5 &4.3&2.2& +38 &10.3 &.09 &120&36  \\
 2134+004 &2&8& A & 0.5 & 0.3 &2.1&1.2&+135&2485  &110&18000&440  \\
      CJ &&& B & 1.0 & 0.8&4.1&3.3&+41 &2503 &21&3900&210  \\
\noalign{\smallskip}
\hline
\noalign{\smallskip}
\end{tabular}
\end{flushleft}
\end{table*}

\subsection{The spectral indices}

The direct comparison between the components found at 5 GHz (paper I)
and at 15 GHz is correct only for the components appearing
well isolated and compact at both frequencies. Otherwise, the
very different UV coverage and spatial resolution introduce
a serious bias in the determination of the spectral index distribution.
Components which are well resolved at 15 GHz might result in spectral
indices artificially steeper than their intrinsic value, due to lower
sensitivity to component sizes of the order of a couple of tens
of mas, or larger.

In addition, we are interested in locating the core in our objects
which is presumed to be compact at both
frequencies and have a flat or inverted spectrum.
In order to find the cores in our objects we determined the spectral
index for isolated and compact components using their fitted
flux densities. When this could not be done,
we degraded the resolution of the 15 GHz image to allow a direct comparison
between the 5 and 15 GHz data, we identified common features
which permitted us to align the VLBI images (which do not have absolute 
astrometry), and we made a spectral index image
to locate the region with the flattest spectral index.

We display images showing the spectral index between 5 and
15 GHz as a gray scale superimposed on the contours of the 15 GHz 
resolution-degraded images,
where darker means flatter. This will emphasize the presence 
of flat components and make it easier to locate the core
(should it be present).

In the 15 GHz data, we generally find fewer extended features
than in our previous 5 GHz images. This is due to the
combined effects of (1) the steep spectral index of the extended
components which would require a much better sensitivity
at 15 GHz than at 5 GHz (while it is only slightly better) 
and (2) the lower sensitivity to extended emission caused by a 
larger gap in the central region of the UV coverage at 15 GHz 
with respect to the 5 GHz data.

In core-jet quasars, the jet is often very short and it is not 
possible to clearly resolve them into individual knots
and/or to determine their positions. In these cases, the image 
registration is not very accurate.

\subsection{The polarization images }

In addition to  total intensity images, we present radio polarization
data in the form of vectors whose length is proportional to the fractional
polarization, and whose direction indicates the position angle of the 
E-field. At 15 GHz,
these orientations are within a few degrees of the intrinsic ones
for Rotation Measures below 100 rad/m$^2$. Furthermore,
when we comment on the E or B field structure, we assume that
differential Faraday Rotation is negligible across the sources,
although this might not be the case. 
We assume that bandwidth depolarization is negligible in our observations
because even an RM of 10$^4$ rad/m$^2$ would produce only a one degree position 
angle rotation across the 16 MHz bandpass at 15 GHz. 

Linearly polarized emission is detected in all the sources
observed, at various percentage levels, depending on
the total intensity of components. 
We cannot detect polarized emission in weak components because of 
the lack of adequate sensitivity. However, even in the brightest components, 
the VLBI 15 GHz data still reveal a rather low fractional polarization - 
a few percent at most. 

The highest frequency at which reliable polarization
data are available from the VLA (Stanghellini et al. 1998)
is 8.4 GHz. We find more polarized emission in our 15 GHz
VLBI data as expected if the (lower frequency and lower resolution) 
VLA data are affected to the combined effects  of beam and Faraday
depolarization. In fact, beam depolarization should be present
in the VLA data due to the gradients in the polarization position
angle seen in the VLBI data.

\subsection{Comments on individual sources}

{\bf 0500+019}: we consider the radio source to be  associated with 
a galaxy at $z=0.583$ rather than a background quasar  
(cf. Stickel et al. 1996a, and see also paper I for references on 
individual objects).  The radio source is S-shaped as is sometimes 
found among GPS radio galaxies, with the northern part 
brighter than the southern part. Polarized emission
is detected at a low fractional level ($\sim 0.3\%$ at 
the total intensity peak position in p.a. $+101$) only in the
brightest component. The VLA total fractional polarization 
at 5 GHz is below 0.1 $\%$ (the reference for total
fractional polarization here and for the other objects
is Stanghellini et al. 1998).
The spectral index image shows that a clear flat-spectrum 
region is not present, the flattest spectral index being
close to the brightest component B. The low resolution
image does not permit us to separate the contribution
of the subcomponents B1 and B2, but we see slightly
darker shades just south of the peak intensity on the low
resolution image, and we suspect the flattest component
is component B2. Therefore, if
the core is located in component B2 or close to it,
this radio source would be a genuine CSO.
This classification is in agreement with the MEM image
showing compact subcomponents corresponding
to B1 and B2, and extended regions at the edges of the
radio source.

{\bf 0738+313}: our 15 GHz VLBA image is dominated by two strong
components. The diffuse emission originating from the southern one
and leading towards the SE seen in the 5 GHz image is still present
in the 15 GHz image, but is less significant.
Component A is weakly polarized (0.5$\%$ at the peak position)
and has an inverted spectrum, thus it is (or it harbors) the core 
of the radio source.  
Component B is 1.5$\%$ polarized at the peak position 
and shows a higher percentage of polarized emission in its boundaries.
The magnetic field structure follows the jet orientation,
running parallel to the local jet axis from component A to B,
then it becomes perpendicular to the source structure in component B where
the radio source bends to the SE. The total fractional
polarization from VLA observations at 8.4 GHz is 3$\%$.
The identification of the core makes this object a core-jet radio source.
This is also confirmed by the MEM image, with 
a jet leading to the South and then suddenly bending to the East.

This radio source is known to have an arc-second scale double (or rather
triple) morphology, with two weak and diffuse lobes on each
side of the compact much stronger component (Stanghellini et al 1998). 
Extended emission around GPS radio sources is 
explained as the effect of a reborn or disrupted radio
source i.e., a source where a new radio activity is in progress while
the relic of the past activity is still present, or a radio
source where a sudden increase of the medium density along
the path of preexisting jets has interrupted the energy supply
to the external lobes (Baum et al. 1990). This explanation can be invoked also
for 0738+313 but the fact that we see a core-jet morphology
at mas resolution favors the hypothesis that this object 
is indeed a classical double with a strong boosting of the inner 
part of the jet.

{\bf 0742+103}: the identification of this radio source
is uncertain --  Stickel et al. (1996b) suggest it is a quasar of
magnitude $m_R\sim 23.6$.
The 15 GHz image suggests a structure reminiscent of 3C48
although the scale is about 100 times smaller. 
The MEM image shows a jet  
leading to NW, then bending to NE and finally widening and 
bending to the East as confirmed by the 5 GHz image.
The spectral index image fails to reveal any
flat region which would allow us to locate the core.

Faint polarized
emission is detected only on the northern tip, where the jet
broadens. The total fractional polarization at 8.4 GHz from
VLA data is below 0.1$\%$.

{\bf 0743-006}: the candidate core component visible
in the 5 GHz image is seen as an elongated structure
in the 15 GHz VLBA image, with just a hint of the 
5 GHz NE jet. The MEM image reveals the elongated
component as a triple structure (a knotty jet?)
and a least-squares fit to the "cleaned and restored'' image 
provides a good fit with 3 components, with slightly
increasing sizes from the southwestern to the
northeastern one, suggesting (taking into consideration
the morphology seen in the 5 GHz image) the presence of an expanding jet.

The low resolution of the 5 GHz image does not allow us
to disentangle the contribution to the flux density of the
distinct components, therefore it is not possible
to determine if any of the 3 components in the 15 GHz image
has a flat spectral index. In this case image registration
did not provide a useful result and a spectral index distribution
image is not presented.  The integrated spectral index
between component A in the 5 GHz image and components A1+A2+A3
in the 15 GHz image is $\alpha = -0.37$. 

The radio source is polarized with a polarization
percentage of $\sim 1\%$ at the peak position in the total
intensity image with a transverse magnetic field.
The fractional polarization increases towards the NE, also changing
its position angle, with an inferred projected
magnetic field which becomes parallel to the local source axis.
It may be an intrinsic change in position angle, or
it might suggest that the core is located at the SW edge of
the source: the jet, then, is likely to cross a region
of high RM which rotates the apparent field orientation;
the RM would rapidly decrease with  distance from the core
and the NE edge would show the intrinsic field orientation.
The total fractional polarization at 8.4 GHz from the VLA
is around 1$\%$.

{\bf 1143-245}: the VLBI image at 15 GHz shows a structure 
roughly in the NS direction extending about 5 mas;
both components are resolved by the present observations.

The MEM image reveals a number of further details, 
with a strong quite compact component and a
weak wiggling jet-like structure leading to the South.

The spectral index is not flat in any
region of the source, despite its core-jet morphology.
As in other similar cases, 
a possible explanation is that the source experienced
significant flux density variability between the
2 epochs, or that the true core is embedded in the
brightest component, or that there are errors in the image registration.

The brightest region has significant polarized emission.
At the peak position in the total intensity
image the fractional polarization is
$\sim 1\%$ in p.a. $-14^\circ$. The VLA data 
show a total fractional polarization of $\sim$1.4$\%$
at 8.4 GHz.

{\bf 1345+125}: the VLBA image at 15 GHz shows the same 
morphology seen at 5 GHz. A strong compact component 
and a jet leading SE with some knots, then bending
to the SW, and widening
into a diffuse lobe which, in the 15 GHz image, is mostly resolved out.

At 15 GHz, we do not detect the weak emission north
of component A, which is present in the 5 GHz image.
Here, the image registration is relatively easy
thanks to the presence of a number of rather isolated, optically thin, 
knots in the jet.

Component A is the flattest one
with $\alpha \sim -0.3$, while component B has  $\alpha \sim -0.5$.
Component A is therefore likely to contain the core.
The northern side is barely visible at 5 GHz,
and while we can assume the jet is relatively beamed
towards us making the counter-jet undetectable, it is puzzling that
we do not see the counter-lobe. 

Polarized emission is detected in the brightest knots
with percentage values of $\sim 1.5\%$ in component A, 
1$\%$ in component B, at the total intensity image
peak positions. Total fractional polarization at 8.4 GHz
is not detected above 0.1$\%$, therefore it is likely
that the VLA 8.4 GHz measurement is affected by intrinsic
and/or beam depolarization.

Stanghellini et al. (in preparation) find  very weak extended
emission on the arc-second scale at 1.4 GHz, around the compact radio source,
with a morphology similar to the parsec scale structure.

{\bf 1404+286}: in  agreement with the
5 GHz image (Stanghellini et al. 1997), the 15 GHz VLBA image shows a rather
symmetric morphology with respect to the location
of the components, but with a large ratio in flux density
between the 2 ``sides" of the radio source.

We detect significant radio emission at the
center of the whole structure at 15 GHz, as already seen
in Kellermann et al. (1998).
Due to the limitation
in the SNR of the 5 GHz image, the spectral index distribution in
the central region could not be determined, but it stands out 
that this region must have an inverted spectrum, as also confirmed by
multi-frequency VLBA observations (Stanghellini et al, in preparation).
On the basis of the symmetry in the radio structure, the central weak
region (Fig. 8a) is likely to harbor the core and possibly two short jets.
The MEM image shows an extended feature at the NE edge
and several compact knots very well aligned along
the source axis.

Polarized emission is detected at the boundaries of the stronger component,
at a local level of several percent, while VLA observations at 8.4 GHz
do not detect any polarized emission above 0.2$\%$. 
Gallimore et al.  (1999) do not detect the 21 cm HI line in absorption 
against the radio source. However, 
Kameno et al. (2000) suggest that the source is free-free absorbed by 
an ionized medium with electron density $n_e \sim 600 - 7\times 10^5$ 
cm$^{-3}$. This medium would be capable of depolarizing the radio source.

{\bf 2126-158}: our VLBI image shows an elongated NS 
core-jet structure with a total extension of about 3 mas. 
The comparison with the 5 GHz image reveals that the
peak of the emission has a flat spectrum, and we identify it
as the core.

The MEM image shows a weak jet emerging from the
core and leading with a curved trajectory to the South.
The spot on the opposite side of the core may be
an artifact but the 5 GHz image also shows 
a hint of emission in that direction, so it is also possible
in this source we detect traces of a counter-jet. 

The core region is $\sim 1 \%$ polarized at p.a. $+52^\circ$,
while the total fractional polarization at 8.4 GHz is
below 0.2$\%$.

{\bf 2128+048}: the 15 GHz VLBA image shows an aligned structure
with four main components at approximately P.A. $-30^\circ$.
The outer components are more resolved than the inner ones
though the southernmost is rather weak.
 
The comparison
with the 5 GHz image indicates that there are no really flat spectrum
components.  Component B seems to have the flattest 
spectral index region reaching $-0.4$, but this may be caused by  poor
alignment (which has been done using component A) between 5 and 15 GHz.
The spectral index using the integrated flux densities
does not provide any further indication of the location of the core,
with $\alpha \sim -0.9$ for component A, and $\sim -1.2$ for component B, 
suggesting they are likely knots in a jet.
However, based on morphology we can reasonably identify the outer
components as the hot spots, and the radio source as
a CSO. 

Polarized emission is detected above the 5 $\sigma$
r.m.s. noise only in the northern component (C$+$D) at a $\sim 5 \%$ level
and with an inferred magnetic field perpendicular (E-vector P.A. $-37^\circ$) to the 
source axis as expected in a typical hot-spot, where the
magnetic field lines are compressed against the shock front. 
VLA observations at 8.4 GHz do not detect any polarized emission above
0.2$\%$.

{\bf 2134+004}: the VLBI image at 15 GHz 
has a double structure just a few resolution elements across the source.
The least-squared fit on the components
suggests that the eastern component is more
compact, in agreement with the MEM image,
which also suggests another weak middle component.
Due to the strong variability of radio flux density for this object
(see references in Stanghellini et al 1998)
we tentatively classify this source
as a core-jet source with the core being
the more compact eastern component.  
The VLBA image at 8.4 GHz shown by Fey et al. (1996)
is consistent with our view. 

Polarization emission is detected in both
components at a level of $2-3\%$, while
the total fractional polarization at 8.4 GHz
is 0.5$\%$.
The B field has a rather complicated structure
and indicates that beam depolarization may still play a role even
at this frequency/resolution.

\section{Discussion}
\subsection{Polarization}
Our 15 GHz VLBI observations detect local low polarization in
the GPS sources - at most a few percent and typically
less than one percent. Thus, it  seems that GPS radio 
sources on the parsec scale are characterized by 
intrinsically low fractional polarization. 

The low polarization could be due to high Faraday rotation measures
if the sources are interacting with, ionizing, and entraining 
a dense ambient medium (e.g., O'Dea, Baum \& Stanghellini 1991;
Bicknell, Dopita \& O'Dea 1997; Begelman 1999). Another
possibility is that there is a foreground screen not directly
associated with the radio source, e.g., a disk (as suggested
for 0108+388 - Marr, Taylor \& Crawford 2000).  

Alternatively, the low polarization might imply the presence of a 
very tangled magnetic field structure which is not resolved by these data. 
We note that the GPS radio sources are less polarized than the extended (tens of
kpc) radio sources, even for a similar number of resolution elements across 
the source. This suggests that the scale size of the magnetic field
inhomogeneities is a smaller fraction of the source size in the GPS
sources and does not scale linearly as the sources expand. 

\subsection{Cores}
We have combined our 5 and 15 GHz observations to
determine the spectral index structure of the sources and one of the
outstanding issues has been the properties of the ``cores" of the
GPS sources. These are faint even in several quasars, and thus
are hard to characterize. 
The cores in the GPS galaxies are very difficult to locate
either because they are too faint to be clearly detected,
or because they do not have a very flat spectrum.

In general, it seems difficult to detect the cores in GPS radio galaxies
and other CSOs: Peck and Taylor (2000) in a sample of 21 CSO or CSO candidate
were able to identify a core in only 6 objects, and 2 of the
6 components identified as cores have a spectral index steeper than -0.5.
We also note that Snellen et al. (2000) in their sample of faint
GPS radio sources detect cores in only 3 CSOs,  while in 11 objects they
detect only two steep spectrum components and classify these objects 
Compact Doubles (CD). 

We conclude that the cores contribute in general only a moderate amount 
to the source luminosity. This is implied also by the high frequency total 
spectra which allow only a limited amount of flux density from a flat
or inverted spectrum  component. 

The core flux 
density can be predicted to be a few or a fraction of percent for quasars 
and galaxies respectively, if they follow the well known P$_{core}$-P$_{tot}$
relation found for extended, steep spectrum powerful radio sources
(e.g., Giovannini et al. 1988; Zirbel \& Baum 1995; 
see also Fanti et al. 1990 for the core fractional 
luminosity in CSS). 
Thus, although we do not find cores in some of the symmetric sources,
we nevertheless classify them as CSOs based on their overall morphology,
e.g., the presence of features which appear to be two lobes on either
side of the source's (invisible) core. 

Also perhaps surprisingly, the cores in some quasars tend to
be fairly weak. In the images of several GPS quasars shown here
the core does  not dominate the radio emission. 

In only  two sources (0738+313 and 2126-158) do we find a clear case of flat or 
inverted spectrum components, which we identify with the source core. In both 
cases the source has a one-sided core-jet structure.
In the other sources we find at most a flattish spectrum component( -0.5$<\alpha<$-0.3) 
which might contain the core embedded in unresolved transparent spectrum
emission. In a couple of cases, the  evidence for the core is based on compactness
of the candidate component (1143-245, 2134+004).
In the case of 0742+103 and  0743-006 the radio
emission seems dominated by a one-sided jet, with the core
emission absent or just a small fraction of the total flux density.

The low flux density variability of GPS radio sources,
quasars included (with some exceptions, see Stanghellini et al. 1998)
is consistent with the absence or weakness of the cores, where
the flux density variability is expected to occur.

\subsection{Morphology}
In paper I and II we have shown that GPS radio galaxies 
preferentially tend to have symmetric pc-scale radio morphologies (i.e.,
they are CSOs) while the GPS quasars seem to typically have core-jet
structures. 
GPS quasar and galaxies still show different morphologies, but
the higher resolution of the 15 GHz images, together with 
the spectral information makes the classification of
the GPS quasars as core-jet radio sources somewhat too simplistic, 
since the cores are often hidden or missing, and we only detect the presence
of a more or less complex jet-like structure.

The GPS quasars could be those objects where the radio emission
is not dominated by the core alone, but exhibit a significant amount
of emission from one or a few knots in a moderately beamed jet.
These knots may correspond to the place of a shock front where
the electron are re-accelerated (possibly 0742+103, 0743-006)
or a place in an helical jet
where the bulk flow is directed toward the observer (possibly 0738+313)
enhancing its brightness by moderate relativistic beaming
(the highest brightness temperature found is of the
order of $10^{11}$ in 2134+004, below the limit of $10^{12}$ for 
un-beamed objects - cf. Readhead 1994).

We found inner components which have somewhat flatter spectra
than the rest of the radio structure. These might indicate the locations
of the cores. The modest flattening of the spectra could be due to 
blending of the cores with nearby steeper spectrum "jet" material 
because of our resolution, which is rather modest in relation to
the total source size for most objects.

Other explanations are also possible.
a) The characteristics of GPS quasars may depend just on
a particular geometric configuration which  partially or completely
obscures the flat spectrum core, leaving the emission from a 
particular region of the jet as the dominant contribution.
b) It is possible that the cores are obscured in the radio by free-free 
or synchrotron self-absorption. c) It could be that the cores
in the GPS sources are just not as flat as seen in the more strongly
beamed radio source. d) It is possible we do not see the
"true" core because it is beamed away from us; indeed this could partly be
a selection effect due to the optically thin spectral index, which favors
QSOs oriented at relatively large angles.

Here we suggest a possible scenario based on the
observed phenomenology within the context of the current Unified
Scheme (Browne 1983; Antonucci \& Ulvestad 1985; Urry \& Padovani 1995). 
It seems likely that the objects in which the 
cores are not beamed towards us will appear as GPS galaxies, while 
those which are moderately beamed will appear to be GPS quasars, while those
which are very strongly beamed should be flat spectrum
radio sources and will not be identified as GPS at all. 
This has two implications. (1) We will not identify some compact
and possibly young quasars as GPS if they are strongly beamed. 
However the number of those which are strongly beamed is a very small fraction
of the population. (2) The presence of beaming in the quasars will influence 
the redshift distribution and cause it to be different than that
of the GPS galaxies (e.g., O'Dea 1998; Snellen et al. 1998, 1999).

\subsection{Wiggles in the Jets and Binary Black Holes}

Roughly 5-10\% of the GPS galaxies in the complete sample show evidence 
for wiggles or S-shaped symmetry in their radio jets (e.g., 0108+388, 
0500+019, 1345+125, 2210+016, and 2352+495).
This might be due to precession of the radio axis 
due to the influence of a second Black Hole (BH). We now believe that 
massive BHs are present in essentially all normal galaxies (e.g., Ferrarese 
\& Merritt 2000). We also believe
that nuclear activity is triggered through galaxy interaction and merging 
(e.g., Heckman et al. 1986). When galaxies merge, their BHs will  form a 
relatively long lived binary with separation in the range $\sim 0.01 - 1$ 
pc (e.g., Begelman, Blandford \& Rees 1980; Milosavlejic \& Merritt 2001). 
The second BH can influence the orientation of the radio jet and induce
wiggles through (1) orbital motion of the active BH (e.g., Roos, Kaastra 
\& Hummel 1993), (2) geodetic precession of the active BH spin axis 
(Roos et al. 1993), or (3) tidally induced precession of the accretion 
disk of the active BH (Romero et al.  2000).

For example, 0500+019 shows a wiggle with a wavelength $\lambda \sim 8$ mas 
($\sim 30$ pc).
Assuming the jet outflow is relativistic, the period is $P \sim 100$ yr
which is roughly consistent with geodetic precession of the active BH spin 
axis. 

\section{Conclusions}

We present VLBA observations at 15 GHz of 10 GPS radio sources. 

We detect very low polarization in the GPS sources --
i.e., typically less than a few percent and often 
less than one percent. This establishes that low polarization
 of the parsec scale structure is an important defining
characteristic of the GPS sources. 

We find that in several quasars the cores do not dominate
the radio emission even at this relatively high frequency
and/or do not have a particularly flat spectrum. 
The cores are difficult to detect in the
radio galaxies and may also have an unusually rather steep radio spectrum.
Alternately the lack of dominant cores could be due to the fact the
the  GPS quasars are at intermediate angles to the line of sight and
so are only moderately beamed. We suggest that if the quasars were
in fact strongly beamed, they would not be identified as GPS sources. 

These observations confirm a dichotomy in the morphological 
classification, as galaxies tend to be symmetric while quasars are not
-- with the presence of cores confirmed in only some sources previously
considered to be simple core-jet radio sources.

\begin{acknowledgements}
C.S. wishes to thank the STScI Collaborative Visitor Program 
for providing support for his visits.
We have made use of the NASA/IPAC Extragalactic Database, operated by the
Jet Propulsion Laboratory, California Institute of Technology,
under contract with NASA.
The National Radio Astronomy Observatory is a facility of the National 
Science Foundation operated under cooperative agreement by Associated 
Universities, Inc. 
\end{acknowledgements}

\clearpage

\begin{figure*}
      \includegraphics[width=18cm]{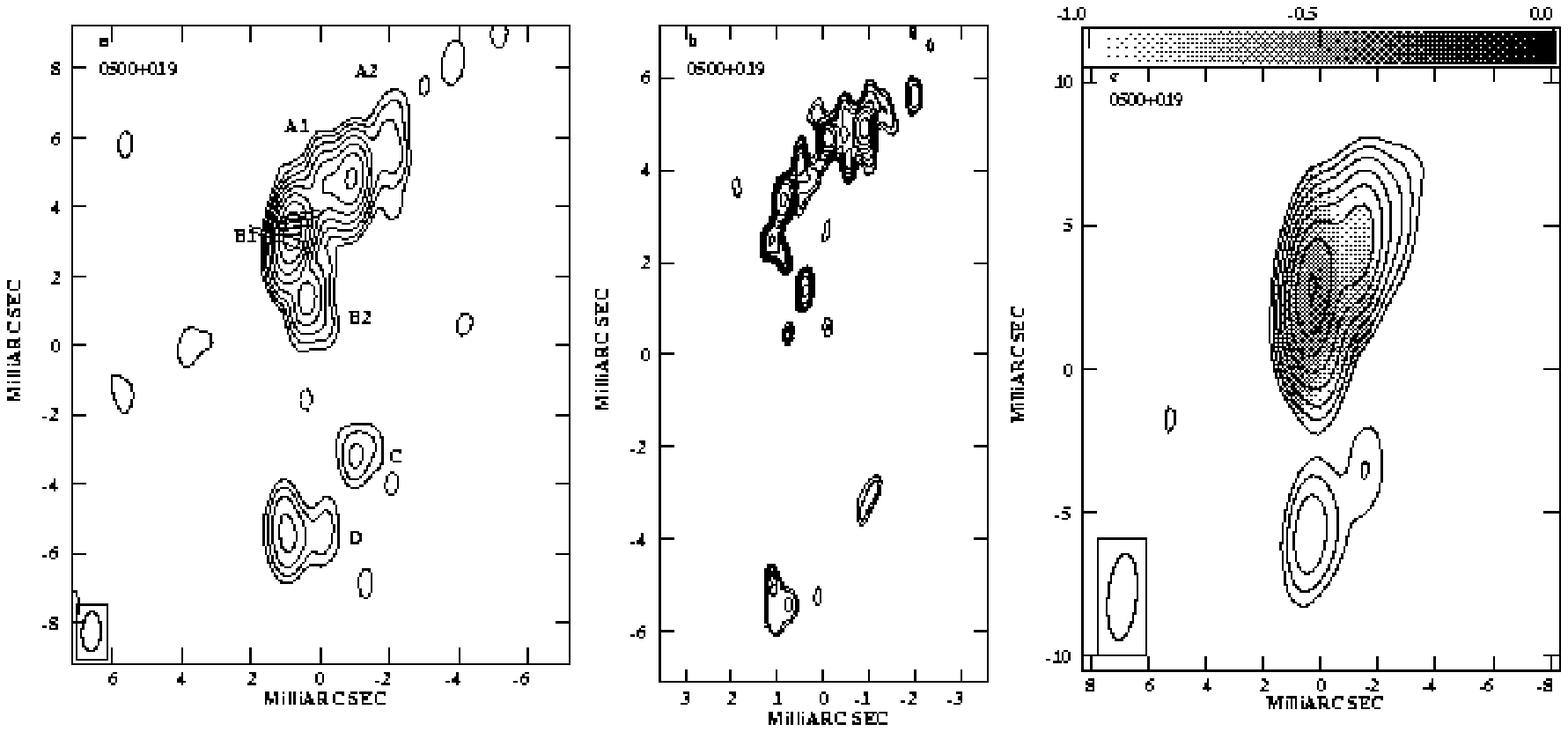}
\caption[]{0500+019: {\bf a)} Contour plot of 15 GHz VLBA image:
the restoring beam is $1.1\times 0.5$ mas
in p.a. $-3^\circ$, the r.m.s. noise on the image is 0.5 mJy/beam,
the peak flux density is 334  mJy/beam. The length of the E-vectors
are proportional to the fractional polarization: 1 mas = 0.005.
{\bf b)} Contour plot of the 15 GHz VLBA image
with maximum entropy deconvolution. {\bf c)} Contour plot of 15 GHz VLBA image: the
restoring beam is $3\times 1$ mas
in p.a. $-6^\circ$, the r.m.s. noise on the image is 1 mJy/beam,
the peak flux density is 425  mJy/beam. The grey scale
displays the spectral index between 5 and 15 GHz, where darker
corresponds to flatter. }
\end{figure*}

\begin{figure*}
      \includegraphics[width=18cm]{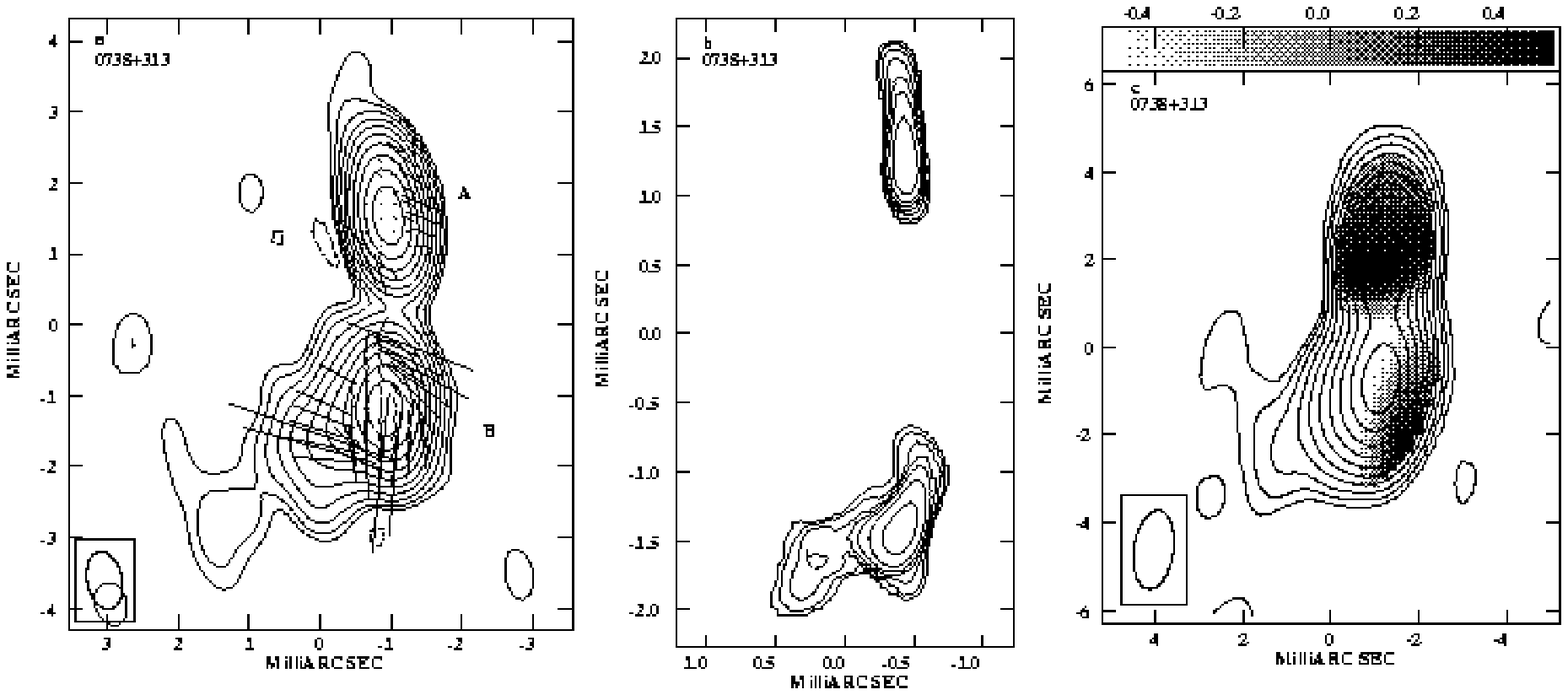}
\caption[]{0738+313: {\bf a)} Contour plot of the 15 GHz VLBA image:
the restoring beam is $0.8\times 0.5$ mas
in p.a. $+12^\circ$, the r.m.s. noise on the image is 0.3 mJy/beam,
the peak flux density is 735  mJy/beam. The length of the E-vectors
is proportional to the fractional polarization: 1 mas = 0.01.
{\bf b)} Contour plot of the 15 GHz VLBA image
with maximum entropy deconvolution. {\bf c)} Contour plot of 15 GHz VLBA image: the
restoring beam is $1.8\times 0.9$ mas
in p.a. $-8^\circ$, the r.m.s. noise on the image is 0.4 mJy/beam,
the peak flux density is 965  mJy/beam. The grey scale
displays the spectral index between 5 and 15 GHz, where darker
corresponds to flatter.}
\end{figure*}

\clearpage

\begin{figure*}
      \includegraphics[width=18cm]{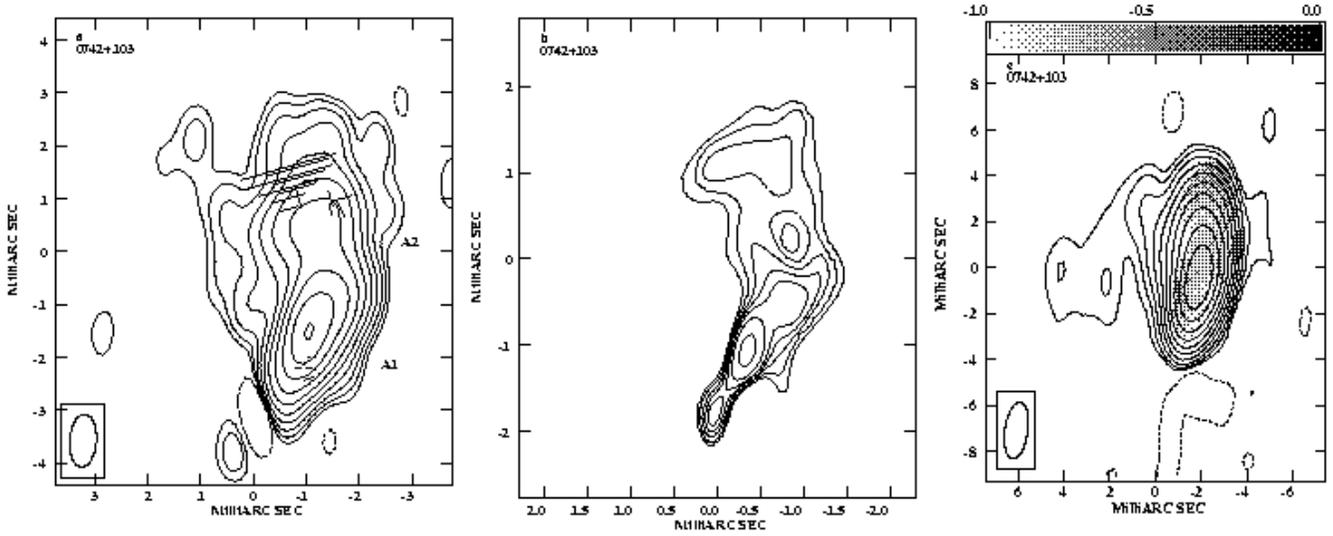}
\caption[]{0742+103: {\bf a)} Contour plot of the 15 GHz VLBA image:
the restoring beam is $1.0\times 0.5$ mas
in p.a. $-4^\circ$, the r.m.s. noise on the image is 0.3 mJy/beam,
the peak flux density is 471  mJy/beam. The length of the E-vectors
correspond to the fractional polarization: 1 mas = 0.05. {\bf b)} Contour plot of the 15 GHz VLBA image
with maximum entropy deconvolution. {\bf c)} Contour plot of 15 GHz VLBA image: the
restoring beam is $2.5\times 1.0$ mas
in p.a. $-8^\circ$, the r.m.s. noise on the image is 0.3 mJy/beam,
the peak flux density is 801  mJy/beam. The grey scale
displays  the spectral index between 5 and 15 GHz, where darker
corresponds to flatter.}
\end{figure*}

\begin{figure*}
      \includegraphics[width=12cm]{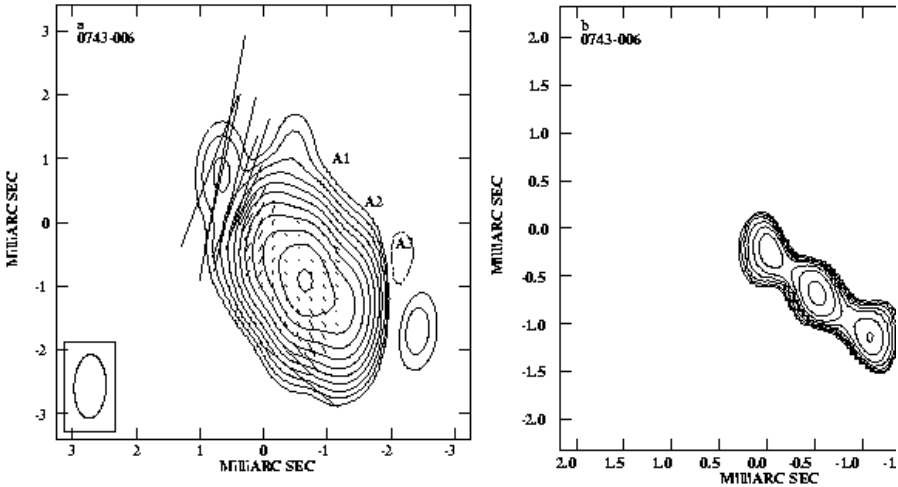}
\caption[]{0743-006: {\bf a)} Contour plot of the 15 GHz VLBA image:
uniform weight, the restoring beam is $1.0\times 0.5$ mas
in p.a. $-2^\circ$, the r.m.s. noise on the image is 0.2 mJy/beam,
the peak flux density is 656  mJy/beam. The length of the E-vectors
is proportional to the fractional polarization: 1 mas = 0.2. {\bf b)}
Contour plot of the 15 GHz VLBA image
with maximum entropy deconvolution.  }
\end{figure*}


\begin{figure*}
      \includegraphics[width=18cm]{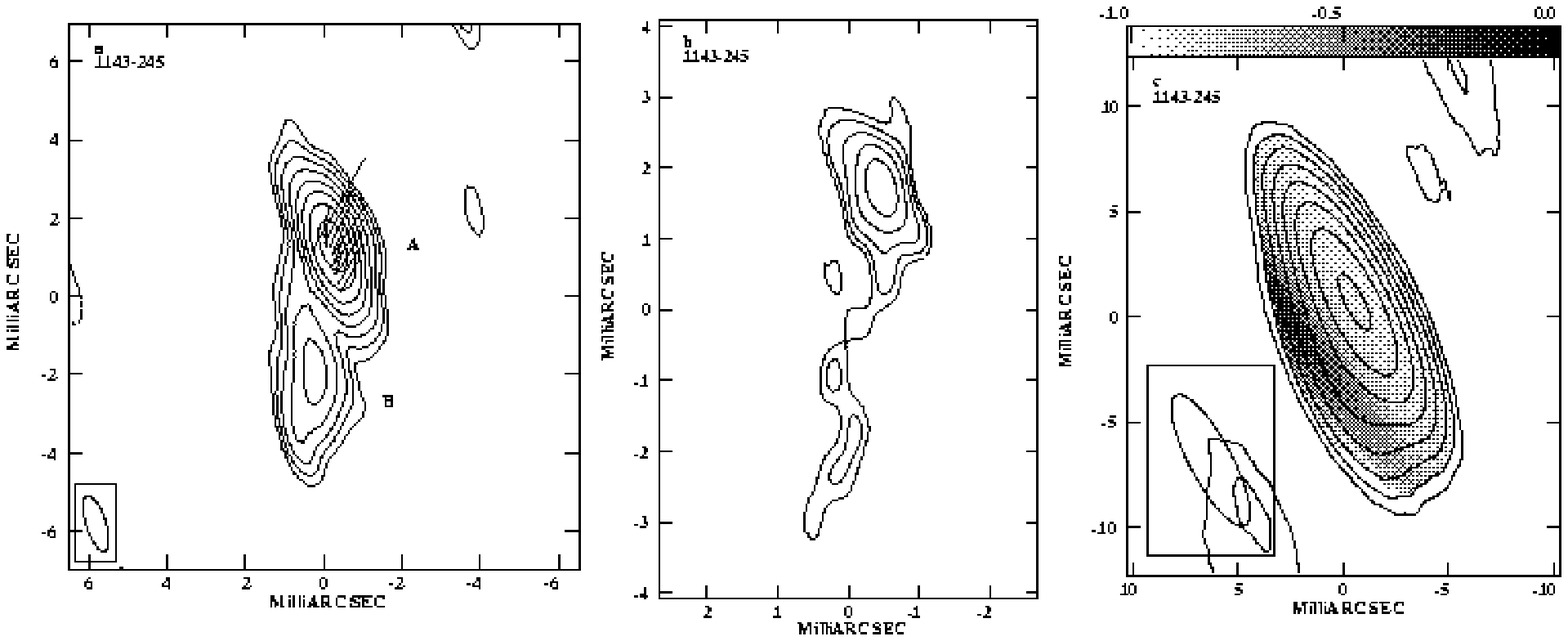}
\caption[]{1143-245: {\bf a)} Contour plot of the 15 GHz VLBA image:
uniform weight, the restoring beam is $1.5\times 0.5$ mas
in p.a. $+18^\circ$, the r.m.s. noise on the image is 0.4 mJy/beam,
the peak flux density is 245  mJy/beam. The length of the E-vectors
is proportional to the fractional polarization: 1 mas = 0.02 {\bf b)} Contour plot of the 15 GHz VLBA image
with maximum entropy deconvolution. {\bf c)} Contour plot of 15 GHz VLBA image: the
restoring beam is $7\times 2$ mas
in p.a. $+28^\circ$, the r.m.s. noise on the image is 1.0 mJy/beam,
the peak flux density is 450  mJy/beam. The grey scale
displays the spectral index between 5 and 15 GHz, where darker
means flatter. }
\end{figure*}


\begin{figure*}
      \includegraphics[width=18cm]{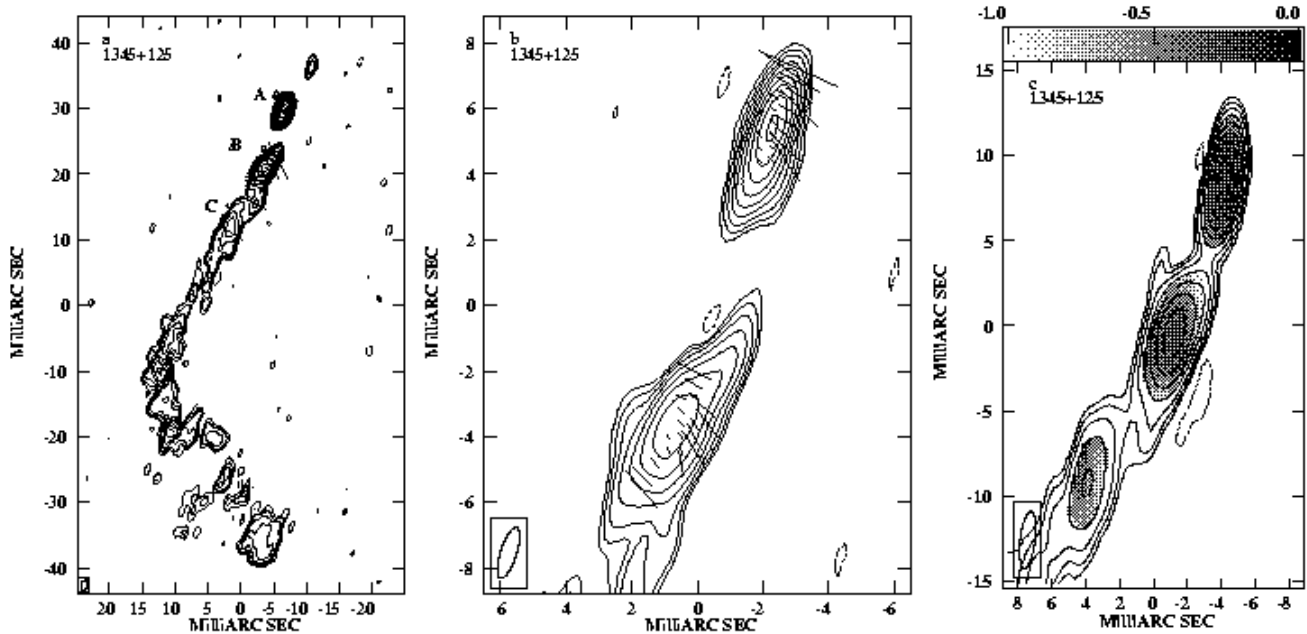}
\caption[]{1345+125: {\bf a)} Contour plot of the 15 GHz VLBA image:
the restoring beam is $1.6\times 0.5$ mas
in p.a. $-18^\circ$, the r.m.s. noise on the image is 0.2 mJy/beam,
the peak flux density is 192  mJy/beam. The length of the E-vectors
correspond to the fractional polarization: 1 mas = 0.05. {\bf b)} Blowup of the previous image.
{\bf c)} Contour plot of 15 GHz VLBA image: the restoring beam is $3.4\times 0.9$ mas
in p.a. $-7^\circ$, the r.m.s. noise on the image is 0.4 mJy/beam,
the peak flux density is 245  mJy/beam. The grey scale
displays the spectral index between 5 and 15 GHz, where darker
means flatter.}
\end{figure*}


\begin{figure*}
      \includegraphics[width=18cm]{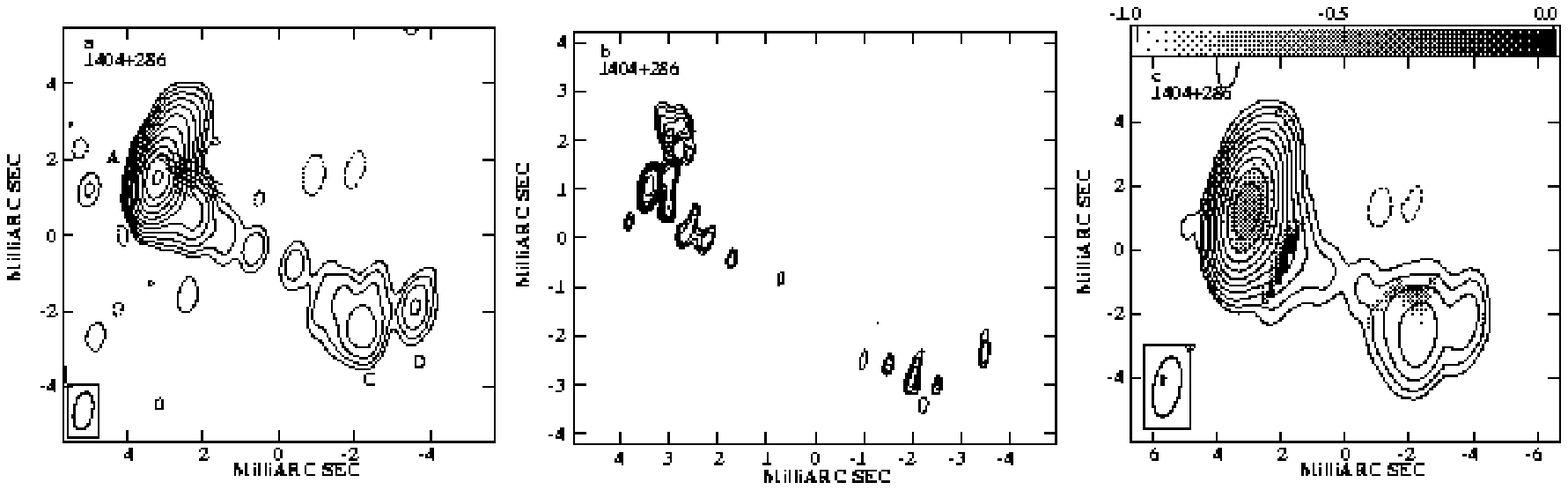}
\caption[]{1404+286: {\bf a)} Contour plot of the 15 GHz VLBA image:
the restoring beam is $1.0\times 0.5$ mas
in p.a. $-11^\circ$, the r.m.s. noise on the image is 0.2 mJy/beam,
the peak flux density is 668  mJy/beam. The length of the E-vectors
are proportional to the fractional polarization: 1 mas = 0.1. {\bf b)}
Contour plot of the 15 GHz VLBA image
with  maximum entropy deconvolution. {\bf c)} Contour plot of 15 GHz VLBA image: the
restoring beam is $1.9\times 0.8$ mas
in p.a. $-11^\circ$, the r.m.s. noise on the image is 0.3 mJy/beam,
the peak flux density is 783 mJy/beam. The grey scale
displays the spectral index between 5 and 15 GHz, where darker
means flatter.}
\end{figure*}


\begin{figure*}
      \includegraphics[width=18cm]{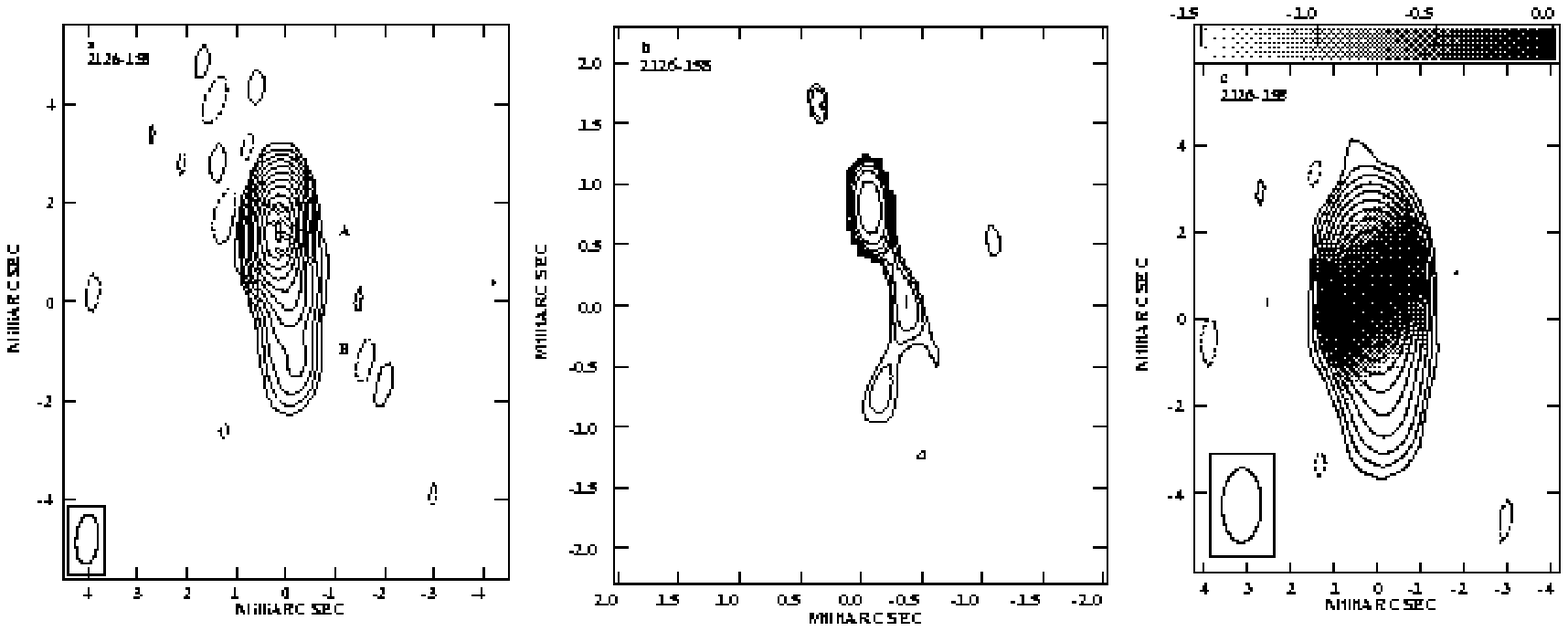}
\caption[]{2126-158: {\bf a)} Contour plot of the 15 GHz VLBA image:
uniform weight, the restoring beam is $1.0\times 0.45$ mas
in p.a. $-4^\circ$, the r.m.s. noise on the image is 0.2 mJy/beam,
the peak flux density is 669  mJy/beam. The length of the E-vectors
is proportional to the fractional polarization: 1 mas = 0.04. {\bf b)}
Contour plot of the 15 GHz VLBA image with maximum
entropy deconvolution. {\bf c)} Contour plot of 15 GHz VLBA image: the
restoring beam is $1.7\times 0.9$ mas
in p.a. $-1^\circ$, the r.m.s. noise on the image is 0.2 mJy/beam,
the peak flux density is 732  mJy/beam. The grey scale
displays  the spectral index between 5 and 15 GHz, where darker
means flatter.}
\end{figure*}


\begin{figure*}
      \includegraphics[width=18cm]{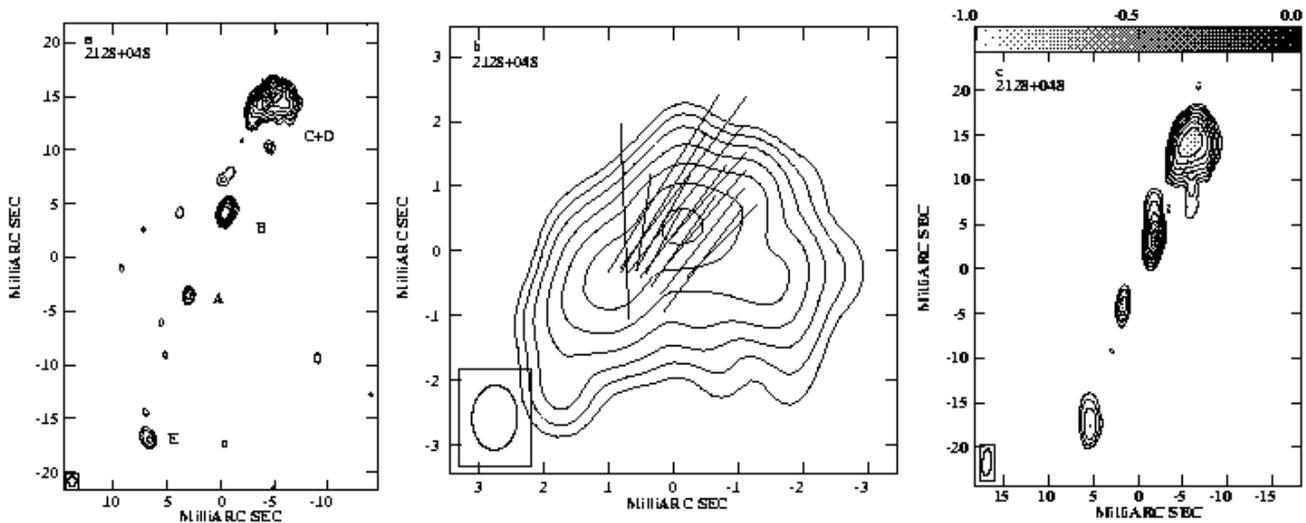}
\caption[]{2128+048: {\bf a)} Contour plot of the 15 GHz VLBA image:
uniform weight, the restoring beam is $1.0\times 0.7$ mas
in p.a. $-2^\circ$, the r.m.s. noise on the image is 0.4 mJy/beam,
the peak flux density is 94  mJy/beam. The length of the E-vectors
is proportional to the fractional polarization: 1 mas = 0.02.
{\bf b)} the northern hot spot, the plot parameters are the
same as for the image of the whole source {\bf c)} Contour plot of 15 GHz VLBA image: the
restoring beam is $3.0\times 1.0$ mas
in p.a. $-5^\circ$, the r.m.s. noise on the image is 0.4 mJy/beam,
the peak flux density is 151  mJy/beam. The grey scale
displays the spectral index between 5 and 15 GHz, where darker
means flatter.}
\end{figure*}


\begin{figure*}
      \includegraphics[width=12cm]{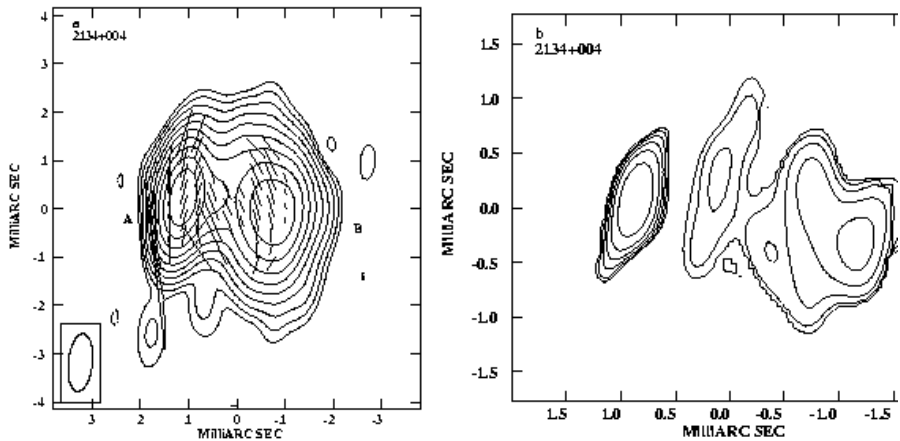}
\caption[]{2134+004: {\bf a)} Contour plot of the 15 GHz VLBA image:
the restoring beam is $1.2\times 0.5$ mas
in p.a. $-6^\circ$, the r.m.s. noise on the image is 0.7 mJy/beam,
the peak flux density is 1938  mJy/beam. The length of the E-vectors
is proportional to the fractional polarization: 1 mas = 0.05. {\bf b)}
Contour plot of the 15 GHz VLBA image with
maximum entropy deconvolution.}
\end{figure*}

\end{document}